\documentclass[journal]{IEEEtran}
\usepackage{cite}

\ifCLASSINFOpdf
\else
\fi
\usepackage{amsmath}
\usepackage{amsfonts}
\usepackage{amssymb}
\usepackage{mathtools}
\usepackage{amsthm}
\usepackage{algorithmic}

\usepackage[linesnumbered,ruled,lined]{algorithm2e}

\usepackage{array}
\usepackage{caption}
\usepackage{subcaption}

\allowdisplaybreaks[4]
\usepackage[subtle,tracking=normal]{savetrees}

\begin{document}
%
\title{Rate-Splitting Multiple Access for Coexistence of Semantic and Bit Communications }
%
%
%

\author{{Yuanwen Liu, Bruno~Clerckx,~\IEEEmembership{Fellow,~IEEE}}
     \thanks{Yuanwen Liu is with the Department of Electrical and Electronic Engineering, Imperial College London, London SW7 2AZ, U.K. (e-mail: y.liu21@imperial.ac.uk).}
 \thanks{Bruno Clerckx is with the Department of Electrical and Electronic Engineering, Imperial College London, London SW7 2AZ, U.K. (e-mail:b.clerckx@imperial.ac.uk).}
 
 \thanks{This work has been partially supported by UKRI grant EP/X040569/1, EP/Y037197/1, EP/X04047X/1, EP/Y037243/1.}}
\maketitle

\begin{abstract}
In the sixth generation (6G) of cellular networks, the demands for capacity and connectivity will increase dramatically to meet the requirements of emerging services for both humans and machines. Semantic communication has shown great potential because of its efficiency, and suitability for users who only care about the semantic meaning. But bit communication is still needed for users requiring original messages. Therefore, there will be a coexistence of semantic and bit communications in future networks. This motivates us to explore how to allocate resources in such a coexistence scenario. We investigate different uplink multiple access (MA) schemes for the coexistence of semantic users and a bit user, namely orthogonal multiple access (OMA), non-orthogonal multiple access (NOMA) and rate-splitting multiple access (RSMA). We characterize the rate regions achieved by those MA schemes. The simulation results show that RSMA always outperforms NOMA and has better performance in high semantic rate regimes compared to OMA. We find that RSMA scheme design, rate region, and power allocation are quite different in the coexistence scenario compared to the bit-only communication, primarily due to the need to consider the understandability in semantic communications. Interestingly, in contrast to bit-only communications where RSMA is capacity achieving without any need for time sharing, in the coexistence scenario, time sharing helps enlarging RSMA rate region.
\end{abstract}


%
\IEEEpeerreviewmaketitle

\section{Introduction}

%
%
%
%

\IEEEPARstart{N}{ew} services and demands for capacity and connectivity pose critical challenges to current networks and call for the development of new technologies in 6G and beyond\cite{9390169,network2030}.

Recently, semantic communication has attracted attention from both academia and industry \cite{ZHANG202260, qin2022semantic,9770094,9679803}. Different from conventional bit communication, semantic communication aims to transmit the semantic meaning of the original messages. In this way, the source data is compressed dramatically while the semantic meaning remains the same \cite{qin2022semantic}. Semantic communication is not a new concept, and it was first proposed in \cite{shannon1949mathematical}. \cite{shannon1949mathematical} categorized communication problems into three levels: \textit{(1) How accurately can the communication symbols be transmitted? (2) How precisely do the transmitted symbols convey the desired meaning? (3) How effectively does the received meaning affect conduct in the desired way?} Semantic communications focus on the second level. Some theoretical works have already been done. \cite{carnap1952outline} gave the concept of the amount of semantic information based on logical probability functions and tentatively defined semantic noise.
\cite{6004632} proposed a model-theoretical framework for semantic data compression and reliable semantic communication. \cite{8476247} proposed a framework of semantic communication, and it obtained the optimal transmission policies which minimized the end-to-end average semantic error based on the Bayesian game.

Although some theoretical works of semantic communication have been done, most of the works in the general area of communications focused on the implementation of the first level of communication. The second level, the semantic problem, was not fully investigated. Thanks to the development of deep learning (DL) based natural language processing (NLP), the implementation of semantic communication is being explored. \cite{8461983} designed a joint source-channel coding of text based on DL for transmitting the meaning of text. \cite{8723589} proposed a joint source and channel code for wireless image transmission, and it directly mapped the image pixel values to the channel input symbols instead of depending on explicit codes for either compression or error detection, and it outperformed the conventional method in low signal-to-noise ratio (SNR) and limited bandwidth region. \cite{9450827} designed a DL-based system to recover the speech signal in semantic communication. \cite{9398576} proposed a DL-based semantic communication system for text transmission to maximize the system capacity and minimize semantic error. \cite{9830752} proposed a semantic communication framework for a multi-user scenario, and it unified the structure of transmitters for different tasks, image retrieval, machine translation, and visual question answering. \cite{10431795} developed a DL-based semantic communication system which can serve many different tasks with multiple modalities of data. 

Semantic has shown its potential for future communication applications, but this does not mean that conventional bit communication will disappear in future networks. A prerequisite of applying semantic communication is that users only care about the meaning of the messages and do not care whether the received messages are the same as the transmitted ones, i.e. some sensors prefer transmitting semantic messages such as significant status changes rather than raw data. However, some users will still require original messages, i.e. some human users collect raw data for detailed analysis. Therefore, future networks should accommodate both bit users and semantic users. This motivates us to investigate suitable multiple access (MA) schemes for the semantic-user-and-bit-user coexistence scenario. OMA and NOMA are two commonly used MA techniques, and they use different interference management methods \cite{clerckx2024multiple}. OMA allocates isolated resources to users. Although users do not interfere with each other, it is not efficient when the number of users increases. NOMA is more spectral efficient than OMA. Users can share the same resources, and interference is fully decoded by successive interference cancellation (SIC). RSMA is a more general MA technique \cite{9831440,10038476}. In the uplink, each user splits its message into multiple parts and encodes them into multiple streams, and then users transmit the superposition of the streams \cite{485709}. These streams are decoded by SIC at the receiver. Although RSMA also uses SIC as NOMA, the message splitting of RSMA provides more flexibility during the SIC procedure. In the downlink, the base station (BS) splits the message of each user into two parts, a common part and a private part. Then, the BS encodes the common parts of all users into a common stream and encodes each private part into private streams. The BS transmits this superposition of streams to the users, and each user decodes the common stream first and then decodes its private stream \cite{mao2018rate}. 

Numerous works have shown the superiority of RSMA for bit communications. In the downlink, the universality of RSMA is shown in \cite{mao2018rate}. RSMA can softly bridge space-division multiple access (SDMA) and NOMA and outperform them. \cite{9382277} show the flexibility of RSMA, which can adapt to varying network loads and channel directions and strengths. \cite{8000591,9940255} show that RSMA is robust to imperfect channel state information (CSI). \cite{9663192,8846706,9844445,10411856} show that RSMA can enhance spectrum efficiency in various scenarios such as multiple-input-multiple-output (MIMO), unicast, multicast, satellite-terrestrial networks, beyond diagonal reconfigurable intelligent surface. \cite{8846706,9650662} show the advantage of RSMA from an energy efficiency perspective. \cite{9893376,9991090,dizdar2023rsma} show that RSMA can enhance the quality of service (QoS) and user fairness. RSMA also shows its promising performances in the finite blocklength regime \cite{9831048}. Since RSMA can achieve the same rate as SDMA and NOMA with a shorter blocklength, it has the potential to be used in low-latency applications. \cite{9967957,9771854} investigated the tradeoff between spectral efficiency and secrecy performance, and shows that RSMA still has rate benefits while guaranteeing secrecy. \cite{10485496} proposes the low-complexity design for RSMA with considering the practical receivers implementation and shows that RSMA still holds advantages over SDMA. The prototype of downlink RSMA is built and its performance is evaluated \cite{lyu2023ratesplitting} and RSMA shows better performance than SDMA and NOMA. In the uplink, RSMA also shows its advantages in improving spectrum efficiency \cite{9257190,10190330,10102273}, enhancing user fairness \cite{xu2024maxmin, xu2024maxminmimo,10510891,10517305} in MIMO, finite blocklength regime and the coexistence of enhanced mobile-broadband (eMBB) and ultra-reliable low-latency communication (URLLC). \cite{liu2024performance} shows that RSMA with hybrid automatic repeat request has the potential to improve reliability. \cite{10330667} shows that uplink RSMA has the potential to be applied in low latency applications with low complexity.

There are also works investigating downlink RSMA for semantic communication. In \cite{10032275}, the BS encoded common knowledge into the common stream and encoded the information for each user into the private stream, and a problem is formulated to minimize the communication and computation energy consumption. Simulation results showed that RSMA has higher energy efficiency. \cite{10428108} designed a framework of RSMA for semantic-aware networks, and shows that RSMA has a lower Age of Incorrect Information (AoII) compared to SDMA. \cite{10333452,zhao2024probabilistic} investigated RSMA for semantic communication based on probability graphs. Since compression brings additional computational resource consumption, \cite{10333452} aimed to minimize the total consumption and \cite{zhao2024probabilistic} maximized the sum of semantic rates subject to the transmission power, semantic compress ratio and rate allocation constraints. The results show that RSMA is energy efficient and can achieve a higher rate after compression compared to SDMA. \cite{10303275} applied RSMA to a task-oriented semantic information transmission framework for image transmission and showed RSMA improved the quality of experience.

Although MA techniques have been studied widely in the bit communication regime, the design of MA techniques for semantic-user-and-bit-user coexistence scenarios is still in its infancy. \cite{Li_2023} studied the designs of NOMA in the downlink, \cite{9953095,10530540} studied the NOMA designs for the coexistence scenario in the downlink. \cite{10158994} studied the coexistence of one bit user and one semantic user in the uplink. In light of this literature, an interesting question arises: Since RSMA has already shown lots of good properties in bit-communication and in downlink semantic communication, will RSMA still hold its advantages in this coexistence scenario? Our work tackles this question by investigating MA schemes for a semantic-user-and-bit-user coexistence scenario in the uplink, simply referred to as coexistence in the sequel. Unlike bit communications, semantic communications also require understandability, which means that the meaning of received messages should be unchanged. Besides, the definition of transmission efficiency is also different from bit communications. Therefore, these specificities of semantic communications should be considered in the MA scheme designs for a coexistence scenario. This work focuses on the coexistence of multiple semantic users transmitting text and one bit user, and the text transmission framework is based on \cite{9398576}. The contributions of this work are below.
\begin{itemize}
\item{This work investigates different MA schemes for the coexistence scenario in the uplink, namely OMA, NOMA and RSMA. To our best knowledge, it is the first paper studying the MA schemes for the coexistence scenario in the uplink. In the coexistence scenario, besides the requirement of bit users, the transmission efficiency and understandability of semantic users also need to be considered. Therefore, it is necessary to take into account those new factors when designing the MA schemes for the coexistence scenario. Besides commonly used OMA and NOMA, we propose a RSMA scheme for the coexistence scenario.}
\item{We characterize the rate region of these three schemes. While in conventional bit communication, the rate region is the set of bit rate pairs that the transmitters can achieve simultaneously, in this coexistence scenario the rate region is the set of achievable semantic rate and bit rate pairs. To obtain the boundary of the rate regions, we fix the semantic rate and then find the maximum achievable bit rate. For OMA, we consider frequency division multiple access (FDMA) and transform this problem into finding the minimum bandwidth required by semantic users. For NOMA and RSMA, we propose a successive convex approximation (SCA) algorithm to solve the problems. }
\item{We simulate the performances of OMA, NOMA and RSMA. We show the rate regions of these three schemes with different numbers of semantic users and numbers of semantic symbols per word. These results show that in the coexistence scenario, RSMA has a larger rate region than NOMA because of its flexibility in the decoding procedure. Although RSMA cannot outperform OMA in low semantic rate regimes because an interference-free resource allocation strategy is preferable, in most situations RSMA has better performances, especially in high semantic rate regimes. Then we present the relation between the number of semantic users and the achievable bit rate. The superiority of RSMA is more obvious as the number of semantic users increases. We also simulate the rate regions of different sentence similarity thresholds, and we show that the improvement of RSMA is more salient when the sentence similarity threshold is high.}
\item{We summarize the main differences between RSMA for bit communications and RSMA for the coexistence scenario. Firstly, in the RSMA scheme design, our RSMA for coexistence does not split the semantic users and allows the bit user to split multiple times. This contrasts with conventional RSMA for bit communications where all users, except one, are split into two parts. We use this scheme because splitting the semantic user may hinder the understandability and splitting bit users multiple times can provide more flexibility in the decoding procedure. Secondly, we observe that the rate region of RSMA for the coexistence is quite different from the one of bit communications due to the constraints on the understandability and transmission efficiency of semantic users. Then we contrast the power allocation strategies of RSMA for the coexistence and RSMA for bit communications to assess the impact of semantic communications on the RSMA designs further. Finally, we find that in the coexistence RSMA cannot always outperform OMA due to the need to account for the understandability of semantic users, while in bit communications RSMA always performs better than OMA and NOMA. However, we can use time sharing between OMA and RSMA to obtain the largest rate region.
}
\end{itemize}

The organization of this paper is summarised below. Section \ref{sec:2} introduces the architecture of the semantic communication framework, DeepSC \cite{9398576}, and the metric of semantic communication and its approximation. Section \ref{sec:3} presented the OMA, NOMA and RSMA schemes for the coexistence scenario and proposed SCA algorithm. Section \ref{sec:4} demonstrates the numerical results of different MA schemes. Section \ref{sec:5} concludes this work.


\textit{Notations}: We use lower-case and bold-face lower-case letters to represent scalars and vectors, respectively. $\mathbb{C}$ denotes the complex numbers set. 


\section{Semantic communication metric and approximation}\label{sec:2}
In this section, we will give a brief introduction to the architecture of DeepSC \cite{9398576} and a new metric for semantic communication \cite{9763856}, semantic rate. Then, we will introduce a data regression method to approximate the key parameter of semantic text transmission, namely sentence similarity \cite{9953095}.

\begin{figure*}[]
\normalsize


\centering
\includegraphics[scale=0.7]{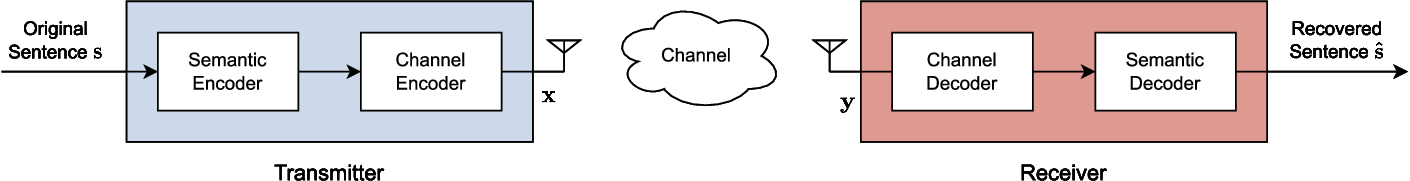}
\caption{The illustration of the framework of DeepSC.}\label{fig:semantic_system_diagram}
\vspace*{4pt}
\end{figure*}

\subsection{Semantic Communication Metric}
DeepSC is a state-of-the-art semantic text transmission framework, and its architecture is shown in Fig. \ref{fig:semantic_system_diagram}. The transmitter of DeepSC has two parts, a semantic encoder and a channel encoder. First, the transmitter extracts the features of the original text $\mathbf{s}=\left[w_1, w_2, ..., w_L\right]$, where $w_l$ is the $l$-th word and $L$ is the average number of words per sentence, and then maps extracted features into the transmitted semantic symbols. Therefore, a word is represented by several semantic symbols, and the transmitted signal is $\mathbf{x} \in \mathbb{C}^{1 \times KL}$, where $K$ is the average number of semantic symbols per word. Considering that both the transmitter and receiver have one antenna, the received signal $\mathbf{y}=h\mathbf{x}+\mathbf{n}$, where $h \in \mathbb{C}^{1 \times 1}$ is the channel coefficient and $\mathbf{n} \in \mathbb{C}^{1 \times KL}$ is the noise. Then the receiver uses channel decoder and semantic decoder to recover the original sentence and the recovered sentence is denoted as $\mathbf{\hat{s}}$. Both the transmitter and the receiver are realized by neural networks, and they are jointly trained to make the meaning of $\mathbf{\hat{s}}$ the same as the meaning of the original sentence $\mathbf{s}$. Thus, a parameter that measures the similarity between the transmitted sentence and the received sentence is important for the system design, and DeepSC uses sentence similarity to measure the similarity. In \cite{9398576}, the sentence similarity is defined as 
\begin{equation}\label{def_semilarity}
\xi \left(\mathbf{\hat{s}}, \mathbf{s} \right)= \frac{\mathbf{\textit{B}_{\Phi}}(\mathbf{s})\cdot \mathbf{\textit{B}_{\Phi}}(\mathbf{\hat{s}})^T}{\Vert\mathbf{\textit{B}_{\Phi}}(\mathbf{s}) \Vert \Vert \mathbf{\textit{B}_{\Phi}}(\mathbf{\hat{s}})^T \Vert},
\end{equation}
where $\mathbf{\textit{B}_{\Phi}}$ is Bidirectional Encoder Representations from Transformers, which is a NLP framework that helps computers understand the meaning of language proposed in \cite{peters2018deep}. $\xi \left(\mathbf{\hat{s}}, \mathbf{s} \right)$ is a value between 0 and 1, and higher $\xi \left(\mathbf{\hat{s}}, \mathbf{s} \right)$ represents higher sentence similarity. The sentence similarity is related to the architecture of the neural networks. For DeepSC, it is decided by signal-to-noise ratio (SNR) $\rho$ of the received signal and $K$ \cite{9398576}. Thus, $\xi \left(\mathbf{\hat{s}}, \mathbf{s} \right)$ can be represented by a function of $\rho$ and $K$, i.e. $\xi \left(\mathbf{\hat{s}}, \mathbf{s} \right)=\varepsilon \left( K, \rho \right)$.

\begin{figure}[]
\centering
\includegraphics[scale=0.5]{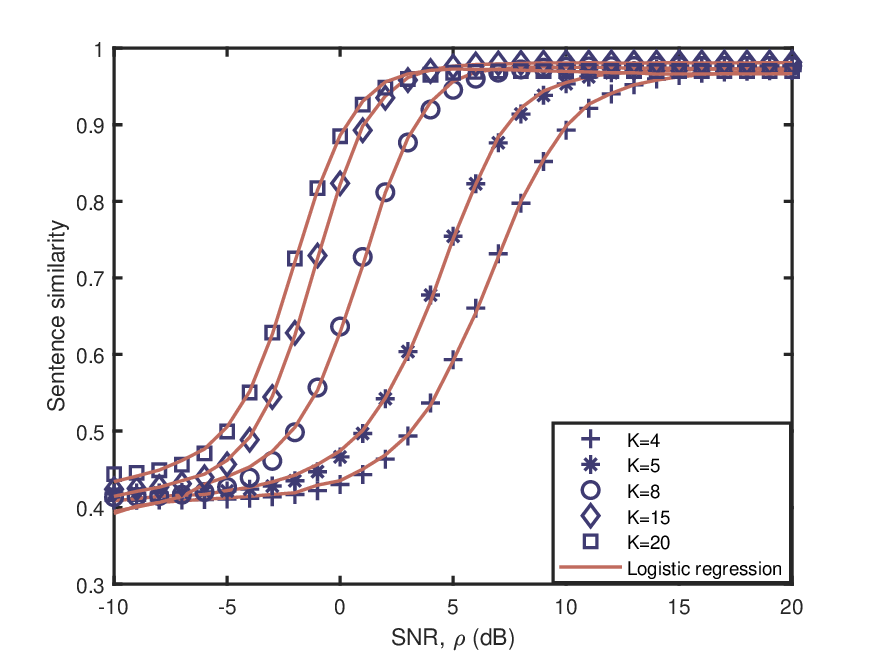}
\caption{The logistic regression for approximating $\varepsilon(\rho,K)$,}\label{fig:SNR_Similarity}
\end{figure}

To further investigate the performance of the semantic communication network in \cite{9398576}, the semantic rate is proposed in \cite{9763856}. Let $I$ denote the average amount of information in the transmitted sentence $\mathbf{s}$, which is measured by the semantic unit ($sut$), so the information contained in each symbol is $\frac{I}{KL} \ (suts/symbol)$. Since the symbol rate is equal to the transmission bandwidth, the amount of transmitted information is $\frac{WI}{KL} \ \ (suts/s)$, where $W$ is the transmission bandwidth. At the receiver side, the received amount of information is 
\begin{equation} \label{semantic_rate}
S=\frac{WI}{KL}\varepsilon \left( K,\rho\right),
\end{equation}
and $S$ can describe the semantic rate $(suts/s)$. $I/L$ is related to the source type, and it is a constant value for a given type of source \cite{9763856}. The value of $\varepsilon \left( K,\rho\right)$ is related to the neural network structure, and we can obtain it with a given $K$ and $\rho$ for DeepSC \cite{9398576}. Part of the $\varepsilon \left( K,\rho\right)$ are presented in Fig. \ref{fig:SNR_Similarity} with blue markers.

\subsection{Data Regression Method For Semantic Rate Approximation}
From the above subsection, we know how to obtain sentence similarity $\varepsilon(K,\rho)$ and semantic rate $S$, and how they relate to the average number of semantic symbols per word, $K$, and the received SNR, $\rho$. However, these relations are not presented mathematically, and this could be a problem for semantic communication design. Thus, \cite{9953095} proposed a data regression method to approximate $\varepsilon(K,\rho)$. 

From the blue marks in Fig. \ref{fig:SNR_Similarity}, we can observe that for a given $K$, $\varepsilon(K,\rho)$ is monotonically non-decreasing as $\rho$ increases, and the derivative of $\varepsilon(K,\rho)$ with respect to $\rho$, $\frac{d}{d\rho}\varepsilon(K,\rho)$, increases first and then decreases. According to the observation, \cite{9953095} used the generalized logistic function to approximate $\varepsilon(K,\rho)$, which is
\begin{equation} \label{approximation}
\varepsilon(K,\rho) \approx \tilde{\varepsilon}_K (\rho) \triangleq A_{k,1}+\frac{A_{k,2}-A_{k,1}}{1+e^{-\left(C_{k,1} \rho +C_{k,2}\right)}}.
\end{equation}
For a given $K$, $A_{k,1}$ and $A_{k,2}$ are the lower asymptote and the upper asymptote, respectively, and $C_{k,1}>0$ denotes the logistic growth rate and $C_{k,2}$ decides the logistic mid-point. Then \cite{9953095} applied the minimum mean square error criterion for fitting this generalized logistic function to get the parameters. Part of the results are shown in Fig. \ref{fig:SNR_Similarity} with red lines, and they can approximate $\varepsilon(K,\rho)$ accurately. This approximation (\ref{approximation}) will help us to design the semantic users and the bit user coexistence schemes. 

Since $\varepsilon(K,\rho)$ is only related to $\rho$ for a given $K$, we omit $K$ in the following sections for convenience, and our discussion is based on a specific $K$.

\section{System Model}\label{sec:3}
\begin{figure}
    \centering \hspace{-0cm}\includegraphics[width=0.55\textwidth]{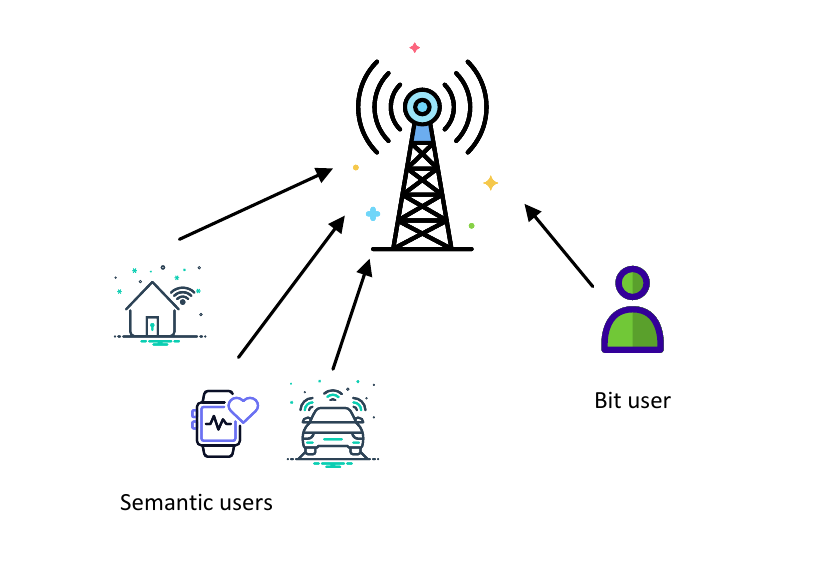}
    \caption{An illustration of semantic users and bit user coexistence.}
    \label{fig:scenario_illustration}
\end{figure}
We consider that there are multiple semantic users and a bit user and they communicate to a common BS. Bit users are interested in the original messages, while semantic users only care about the meaning and most of them are machine-type users. Therefore, the number of semantic users can be much more than the number of bit users, so the scenario that a bit user and multiple semantic users can grasp the essence. An illustration is shown in Fig. \ref{fig:scenario_illustration}. The BS and the users are single-antenna. We use $N_s$ to denote the number of semantic users, $U_{s,j}$ denote the semantic users, where $j\in \mathcal{J}, \mathcal{J}=\{1,2,..., N_s\}$, and $U_b$ denote the bit user. Let $h_{s,j} \in \mathbb{C}^{1 \times 1} $ and $h_b\in \mathbb{C}^{1 \times 1}$ denote the channel coefficients for $U_{s,j}$ and $U_b$, respectively. The transmission power of $U_{s,j}$ and $U_b$ are $p_{s,j}$ and $p_b$, respectively, and the maximum transmission power is $P$ for all users. The total available bandwidth is $w$. Let $x_{s,j}$ denote the normalized semantic symbol of $U_{s,j}$ and $x_b$ denote the normalized conventional symbol of $U_b$. Since multiple users share limited resources, designing an appropriate MA scheme would be important to this coexistence scenario. We first study two commonly used MA techniques, OMA and NOMA. For OMA, here we will consider FDMA. Then we propose a resource allocation scheme relying on RSMA for this coexistence scenario. To evaluate the performances of different MA schemes, we investigate the rate regions of these three schemes. Different from the rate region in conventional bit communication, here the rate region is a set of semantic rate and bit rate pairs. We assume that CSI is known at the BS to study the fundamental performance of the three schemes, and the BS will inform the resource allocation strategies to all the users during the connection setup process. The semantic users require the same service so they should satisfy the same requirements.

\subsection{FDMA}
\begin{figure}[]
\centering
\hspace{0cm}\includegraphics[width=0.5\textwidth]{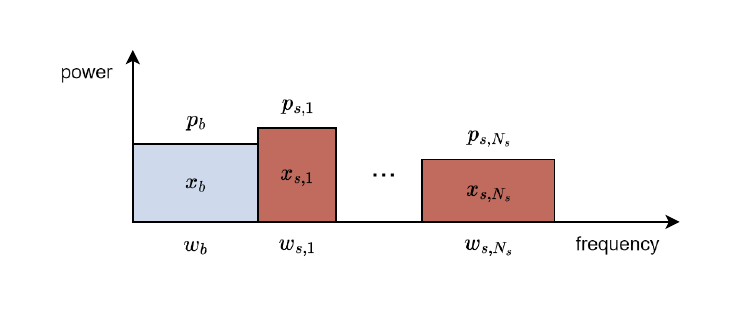}
\caption{The illustration of FDMA.}\label{fig:fdma_allocation}
\end{figure}

In FDMA, the bit user and the semantic users are allocated isolated frequency bands. Let $w_{s,j}$ and $w_b$ denote the bandwidths allocated to $U_{s,j}$ and $U_b$, respectively. The illustration of FDMA is shown in Fig. \ref{fig:fdma_allocation}. The received signal can be represented as 
\begin{equation}
y=\sum_{j=1}^{N_s} h_{s,j} x_{s,j}+ h_b x_b +n,
\end{equation}
where $n$ is the additive white noise. The SINR of the bit user is
\begin{equation}
    \rho_b=\frac{p_b |h_b|^2}{w_b N_0},
\end{equation}
where $N_0$ is the noise spectral power density. The achievable bit rate is 
\begin{equation}
r_b^F=w_b \log_2 \left( 1+\rho_b\right).
\end{equation}
The SINR of $U_{s,j}$ is
\begin{equation}
    \rho_{s,j}=\frac{p_{s,j}|h_{s,j}|^2}{w_{s,j} N_0}.
\end{equation}
According to the  the approximation of sentence similarity in (\ref{approximation}), we have
\begin{equation}
\varepsilon \left( \rho_{s,j} \right)=A_{k,1}+\frac{A_{k,2}-A_{k,1}}{1+e^{-\left(C_{k,1}\rho_{s,j}+C_{k,2} \right)}}.
\end{equation}
The sentence similarity should be higher than a threshold $S^{th}$ so that the semantic message can be decoded \cite{9398576}. $S^{th}$ satisfies $A_{k,1}\leq S^{th} \leq A_{k,2}$ because $A_{k,2}$ and $A_{k,1}$ are the upper bound and the lower bound of the sentence similarity, respectively. Thus, 
\begin{equation}\label{SNR_threshold}
\begin{aligned}
    \varepsilon \left( \rho_{s,j} \right) & \geq S^{th} \\
    \rho_{s,j} & \geq -\frac{1}{C_{k,1}}\left[C_{k,2}+\ln\left(\frac{A_{k,2}-A{k,1}}{S^{th}-A_{k,1}}-1\right)\right]= \gamma_{sem},
\end{aligned}
\end{equation}
which means that the SINR of the semantic users should be higher than $\gamma_{sem}$. Then, the achievable semantic rate for $U_{s,j}$ is
\begin{equation}
S_j^F=\frac{w_jI}{KL}\varepsilon\left(\rho_{s,j}\right).
\end{equation}
Therefore, the rate region of coexistence with FDMA is

\begin{equation}
\begin{aligned}
    C^{F} \triangleq \bigcup \Bigg\{&\left. \left(S^F, R^F\right):   S^F\leq S_j^F, R_b^F \leq r_b^F, \varepsilon(\rho_{s,j})\geq S^{th},  \right.  \\
     & p_b \leq P, p_{s,j} \leq P, w_b+\sum_{j=1}^{N_s} w_{s,j}\leq w, j\in \mathcal{J}  \Bigg\}.
\end{aligned}
\end{equation}

To characterize the rate region, we can fix an achievable semantic rate $S$, and find the highest achievable rate $r_b^F$. In this way, each semantic rate has a corresponding maximum $r_b^F$, so the rate region can be obtained. Therefore, for a given $S^{th}$ and semantic rate $S$, a problem that maximizes the bit rate can be formulated as below:
\begin{subequations} 
\begin{align}
\max_{\mathbf{w_s}, w_b, \mathbf{p_s}, p_b}  &\  w_b \log_2\left(1+\frac{p_b|h_b|^2}{w_b \sigma_n^2}\right),  \label{problem_fdma} \\
 \mathrm{s.t.} \
& \frac{w_{s,j}}{K}\varepsilon \left(\frac{p_{s,j}|h_{s,j}|^2}{w_{s,j}\sigma_n^2}\right) \geq S, \ \forall j\in \mathcal{J},\label{pfdma_1}\\
& \frac{p_{s,j}|h_{s,j}|^2}{w_{s,j}\sigma_n^2} \geq \gamma_{sem},  \ \forall j\in \mathcal{J}, \label{pfdma_2} \\
& p_b\leq P, \label{pfdma_3} \\
& p_{s,j} \leq P, \ \forall j \in \mathcal{J}, \label{pfdma_4} \\
& \sum_{j\in \mathcal{J}} w_{s,j} +w_b \leq w, \label{pfdma_5}
\end{align}
\end{subequations}
where $\mathbf{w_s}=\left[w_{s,1}, w_{s,2}, ..., w_{s,N_s}\right]$, $\mathbf{p_s}=\left[p_{s,1}, p_{s,2}, ..., p_{s,N_s}\right]$. In the optimal solution, $w_b=w-\sum_{j\in\mathcal{J}}w_{s,j}$ and the bit user will use full power to transmit, which means $p_b=P$. The problem can be transformed to
\begin{subequations} 
\begin{align}
\min_{\mathbf{w_s}, \mathbf{p_s}}  &\  \sum_{j\in\mathcal{J}} w_{s,j},  \label{problem_fdma2} \\
 \mathrm{s.t.} \
& (\ref{pfdma_1}), (\ref{pfdma_2}) \ \mathrm{and} \ (\ref{pfdma_4}), \\
& \sum_{j\in\mathcal{J}} w_{s,j}\leq w.
\end{align}
\end{subequations}
To solve this problem, we use the following lemmas.

\textit{Lemma 1:} The semantic rate of $U_{s,j}$ increases when $w_{s,j}$ increases. For a given $S$, the smallest $w_{s,j}$ that satisfies (\ref{pfdma_1}) is $w_{s,j}^{\ast}$ and $\frac{w_{s,j}^{\ast}}{K}\varepsilon \left(\frac{p_{s,j}|h_{s,j}|^2}{w_{s,j}^{\ast}\sigma_n^2}\right) = S$.

\begin{proof}
We use $f(w_{s,j})$ to denote the relation between $w_{s,j}$ and the semantic rate, so
\begin{equation}
\begin{aligned}
&f(w_{s,j})\\
&=\frac{w_{s,j}}{K} \left\{A_{k,1}+\frac{A_{k,2}-A_{k,1}}{1+e^{-\left[10 C_{k,1}\lg\left(\frac{p_{s,j}|h_{s,j}|^2}{w_{s,j}\sigma_n^2}\right)+C_{k,2} \right]}} \right\}.
\end{aligned}
\end{equation}
The partial derivative of $f$ with respect to the $w_{s,j}$ is 
\begin{equation}\label{fdma_partial}
    \begin{aligned}
        &\frac{\partial{f(w_{s,j})}}{\partial{w_{s,j}}}\\
        &=\frac{1}{K} \left[A_{k,1}+\frac{A_{k,2}-A_{k,1}}{1+\left( \frac{p_{s,j}|h_{s,j}|^2}{w_{s,j}\sigma_n^2}\right)^{-\frac{10C_{k,1}}{\ln{10}}}e^{-C_{k,2}} } \right] \\
        &-\frac{1}{K}\left\{(A_{k,2}-A_{k,1})\frac{e^{-C_{k,2}}\frac{10C_{k,1}}{\ln{10}}\left(\frac{p_{s,j}|h_{s,j}|^2}{w_{s,j}\sigma_n^2}\right)^{\frac{-10C_{k,1}}{\ln{10}}}}{\left[1+\left(\frac{p_{s,j}|h_{s,j}|^2}{w_{s,j}\sigma_n^2}\right)^{-\frac{10C_{k,1}}{\ln{10}}}e^{-C_{k,2}}\right]^2} \right\}\\
        &=\frac{1}{K}A_{k,1}\\
        &+\frac{A_{k,2}-A_{k,1}}{K} 
         \frac{1+\left(\frac{p_{s,j}|h_{s,j}|^2}{w_{s,j}\sigma_n^2}\right)^{-\frac{10C_{k,1}}{\ln{10}}}e^{-C_{k,2}}\left(1-\frac{10C_{k,1}}{\ln{10}}\right)}{\left[1+\left(\frac{p_{s,j}|h_{s,j}|^2}{w_{s,j}\sigma_n^2}\right)^{-\frac{10C_{k,1}}{\ln{10}}}e^{-C_{k,2}}\right]^2}.
    \end{aligned}
\end{equation}
The first term and the denominator of the second term are positive in (\ref{fdma_partial}), and we consider the numerator of the second term. $\frac{p_{s,j}|h_{s,j}|^2}{w_{s,j}\sigma_n^2}$ should be no smaller than $\gamma_{sem}$, otherwise, it will violate (\ref{pfdma_2}). Since $A_{k,1}$, $A_{k,2}$, $C_{k,1}$ and $C_{k,2}$ are constant, and specifically $C_{k,1}$ is positive, we can obtain the smallest value of (\ref{fdma_partial}) when $\frac{p_{s,j}|h_{s,j}|^2}{w_{s,j}\sigma_n^2}=\gamma_{sem}$, which is a positive number. Thus, the semantic rate increases when $w_{s,j}$ increases, so the smallest $w_{s,j}$ is $w_{s,j}^{\ast}$ where $\frac{w_{s,j}^{\ast}}{K}\varepsilon \left(\frac{p_{s,j}|h_{s,j}|^2}{w_{s,j}^{\ast}\sigma_n^2}\right) = S$. The proof of Lemma 1 is completed.
\end{proof}
From (\ref{pfdma_1}) and Lemma 1, we can obtain that
\begin{equation}
    w_{s,j}\geq w_{s,j}^{\ast}=w_{s,j}^{l},
\end{equation}
and $w_{s,j}^{\ast}$ satisfies that 
\begin{equation}\label{fdma_rq}
    \frac{w_{s,j}^{\ast}}{K}\varepsilon \left(\frac{p_{s,j}|h_{s,j}|^2}{w_{s,j}^{\ast}\sigma_n^2}\right) = S.
\end{equation}
From (\ref{pfdma_2}), we can obtain that 
\begin{equation}
w_{s,j}\leq \frac{p_{s,j}|h_{s,j}|^2}{\gamma_{sem}\sigma_n^2}=w_{s,j}^{h}. \label{fdma_rq1}
\end{equation}
If (\ref{problem_fdma2}) is feasible, a feasible solution $w_{s,j}$ should be in $\left[w_{s,j}^{l},w_{s,j}^h\right]$ and $w_{s,j}^h \geq w_{s,j}^{l}$. In fact, $w_{s,j}^h$ is always larger than $w_{s,j}^{l}$. (\ref{fdma_rq1}) can be transformed to 
\begin{equation}\label{fdma_rq2}
    p_{s,j}|h_{s,j}|^2=w_{s,j}^h \gamma_{sem} \sigma_n^2,
\end{equation}
and we substitute (\ref{fdma_rq2}) into (\ref{fdma_rq}), we have 
\begin{equation}
    \frac{w_{s,j}^{\ast}}{K}\varepsilon \left(\frac{w_{s,j}^h \gamma_{sem}}{w_{s,j}^{\ast}}\right) = S.
\end{equation}
Therefore, $w_{s,j}^h\geq w_{s,j}^{\ast}=w_{s,j}^l$, otherwise, $\varepsilon \left(\frac{w_{s,j}^h \gamma_{sem}}{w_{s,j}^{\ast}}\right)<S^{th}$ will violate (\ref{pfdma_2}). Since $w_{s,j}^h$ is always larger than $w_{s,j}^l$, the optimal $w_{s,j}$ is $w_{s,j}^l$. Then the next step is finding the optimal $p_{s,j}$ which gives the least $w_{s,j}^l$. To find the optimal $p_{s,j}$, we use the following lemma.

\textit{Lemma 2:} For a given $w_{s,j}^{\ast}$ which satisfies (\ref{fdma_rq}), the optimal $p_{s,j}$ is $P$.

\begin{proof}
When $p_{s,j}=P$, $\frac{w_{s,j}^{\ast}}{K}\varepsilon \left(\frac{P |h_{s,j}|^2}{w_{s,j}^{\ast}\sigma_n^2}\right) = S$. We assume that there exists  $w_{s,j}^{\ast \ast}<w_{s,j}^{\ast}$ and $p_{s,j}^{\ast \ast}<P$, and they satisfy $\frac{w_{s,j}^{\ast \ast}}{K}\varepsilon \left(\frac{p_{s,j}^{\ast \ast}|h_{s,j}|^2}{w_{s,j}^{\ast \ast}\sigma_n^2}\right) = S$ and $\frac{p_{s,j}^{\ast \ast}|h_{s,j}|^2}{w_{s,j}^{\ast \ast}\sigma_n^2}\geq \gamma_{sem}$. Obviously, the semantic rate increases when $p_{s,j}$ increases, so $\frac{w_{s,j}^{\ast \ast}}{K}\varepsilon \left(\frac{p_{s,j}^{\ast \ast}|h_{s,j}|^2}{w_{s,j}^{\ast \ast}\sigma_n^2}\right)< \frac{w_{s,j}^{\ast \ast}}{K}\varepsilon \left(\frac{P |h_{s,j}|^2}{w_{s,j}^{\ast \ast}\sigma_n^2}\right)$. From Lemma 1, we have $\frac{w_{s,j}^{\ast \ast}}{K}\varepsilon \left(\frac{P |h_{s,j}|^2}{w_{s,j}^{\ast \ast}\sigma_n^2}\right)<\frac{w_{s,j}^{\ast }}{K}\varepsilon \left(\frac{P |h_{s,j}|^2}{w_{s,j}^{\ast}\sigma_n^2}\right)=S$, which contradicts to the assumptions. The proof of Lemma 2 is completed.
\end{proof}

Therefore, we have the minimum $\sum_{j \in \mathcal{J}} w_{s,j}$ when $p_{s,j}=P$ and $w_{s,j}=w_{s,j}^{\ast}$ where $w_{s,j}^{\ast}$  satisfies (\ref{fdma_rq}). If $\sum_{j \in \mathcal{J}} w_{s,j} \leq w$, this solution is feasible, otherwise, problem (\ref{problem_fdma2}) is unsolvable. Thus, we find the highest achievable bit rate for every given $S$ and then we can obtain the rate region.

\subsection{NOMA}
\begin{figure}[]
\centering
\includegraphics[width=0.4\textwidth]{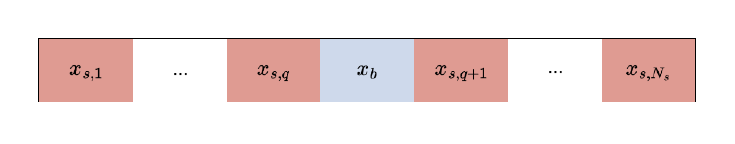}
\caption{The illustration of NOMA.}\label{fig:noma_allocation}
\end{figure}

In NOMA, semantic users and the bit user share the same bandwidth as shown in Fig. \ref{fig:noma_allocation}. The received signal can be represented as
\begin{equation}
y=\sum_{j \in \mathcal{J}} \sqrt{p_{s,j}} h_{s,j} x_{s,j}+\sqrt{p_b} h_b x_b +n,
\end{equation}
Since semantic users and the bit user will use the same band $w$, we use $\sigma_n ^2$ to denote the noise power over $w$ for convenience. BS uses SIC to decode messages, and it is important to design an appropriate decoding order. First, we consider the decoding order among the semantic users. We assume that $U_{s,j}$ is the user with the $j$-th largest channel gain. From (\ref{SNR_threshold}), we know that for a given sentence similarity threshold $S^{th}$, the SINR of the semantic users should achieve $\gamma^{sem}$, so we decode the users with better channel conditions first, and the decoding order of the semantic users is $U_{s,1}, U_{s,2}, ..., U_{s,N_s}$. Therefore, when there are $N_s$ semantic users and a bit user, there are $N_s+1$ decoding orders in total. 

\begin{figure}[]
\centering
\includegraphics[scale=0.6]{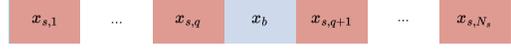}
\caption{The decoding order of NOMA.}\label{fig:noma_decoding}
\end{figure}

We assume that the bit user will be decoded after $q$ semantic users, where $0\leq q\leq N_s$, and $q=0$ means that the bit user will be decoded before all the semantic users. An illustration of the decoding order of NOMA is shown in Fig. \ref{fig:noma_decoding}. The SINR of the bit user is 
\begin{equation}
\rho_b=\frac{p_b |h_b|^2}{\sigma_n^2+\sum_{i=q+1}^{N_s} p_{s,i}|h_{s,i}|^2},
\end{equation}
and the achievable rate is 
\begin{equation}
r_b^N=w\log_2\left(1+\frac{p_b |h_b|^2}{\sigma_n^2+\sum_{i=q+1}^{N_s} p_{s,i}|h_{s,i}|^2}\right). 
\end{equation}
The SINR of $U_{s,j}$ is
\begin{equation}
 \rho_{s,j}=
 \begin{cases}
    \frac{p_{s,j}|h_{s,j}|^2}{\sigma_n^2+\sum_{i=j+1}^{N_s}p_{s,i}|h_{s,i}|^2+p_b |h_b|^2}, \ &j\leq q, \\
    \frac{p_{s,j}|h_{s,j}|^2}{\sigma_n^2+\sum_{i=j+1}^{N_s}p_{s,i}|h_{s,i}|^2}, \ &j>q,
\end{cases}
\end{equation}
and the achievable semantic rate is 
\begin{equation}
S_j^N=\frac{wI}{KL} \varepsilon \left( \rho_{s,j}\right).
\end{equation}
The rate region of NOMA is 
\begin{equation}
\begin{aligned}
    C^{N} \triangleq \bigcup \Bigg\{ &\left. \left(S^N, R_b^N\right):  S^N\leq S_j^N, R_b^N \leq r_b^N, \varepsilon(\rho_{s,j})\geq S^{th}, \right. \\ & j\in \mathcal{J}, p_{s,j}\leq P, p_{s,j}\leq P,  q=\{0,1,2,...N_s \}\Bigg\}.
    \end{aligned}
\end{equation}

Similarly, we fix an achievable $S$, and find the maximum achievable $r_b^N$. A problem maximizing the bit rate is formulated as
\begin{subequations}
    \begin{align}
    \max_{\mathbf{p_s}, p_b, q} \ & \log_2\left(1+\frac{p_b |h_b|^2}{\sigma_n^2+\sum_{i=q+1}^{N_s} p_{s,i}|h_{s,i}|^2}\right) \label{noma_problem00}  \\
    \mathrm{s.t.} \ \ 
    & \rho_{s,j} \geq \gamma_{sem}, \ \forall j \in \mathcal{J},   \label{noma_001}  \\
    & \frac{wI}{KL} \varepsilon\left(\rho_{s,j}\right)\geq S, \ \forall j \in \mathcal{J},  \label{noma_002}\\
    & p_{s,j}\leq P, \ \forall j \in \mathcal{J}, \label{noma_003} \\
    & p_b\leq P, \label{noma_004} \\
    & q=0, 1, ..., N_s.    \label{noma_005}
    \end{align}
\end{subequations}
From (\ref{semantic_rate}) and (\ref{approximation}), we can rewrite (\ref{noma_002}) to
\begin{equation} \label{semantic_rate_sinr}
\rho_{s,j} \geq 
\begin{cases}
\gamma_{sem},  & S\leq \frac{S^{th}I}{KL}, \\
-\frac{1}{C_{k,1}}\left[\ln{\left(\frac{A_{k,2}-A_{k,1}}{\frac{KLS}{I}-A_{k,1}} -1\right)} +C_{k,2}\right], & \mathrm{otherwise}. \end{cases}
\end{equation}
(\ref{semantic_rate_sinr}) shows that the semantic rate constraint is equal to a constraint for $\rho_{s,j}$, which is $\rho_{s,j}$ should be higher than a constant and we use $\gamma$ to denote this constant. Note that the maximum $S$ is $\frac{A_{k,2}I}{KL}$ because $A_{k,2}$ is the upper bound of $\varepsilon(\rho_{s,j})$.

We propose a SCA method to solve the problem (\ref{noma_problem00}). Since (\ref{noma_problem00}) is non-convex due to the non-convex rate expressions, we introduce the following variables, $t$, $\rho_b$ and $\sigma_b$. Then the problem is transformed to 
\begin{subequations}
    \begin{align}
    \max_{\mathbf{p_s}, p_b, \rho_b, \sigma_b, q}  & t  \label{noma_problem0}  \\
    \mathrm{s.t.} \ \ & w\log_2 \left(1+\rho_b \right) \geq t, \label{noma_1}\\
    & \frac{p_b |h_b|^2}{\sigma_b}\geq \rho_b, \label{noma_2} \\
    & \sigma_n^2+\sum_{i=q+1}^{N_s} p_{s,i}|h_{s,i}|^2\leq \sigma_b, \label{noma_3}\\
    & \rho_{s,j} \geq \gamma, \ \forall j \in \mathcal{J},  \label{noma_4}\\
    & (\ref{noma_001}) \ \mathrm{and} \ (\ref{noma_003})-(\ref{noma_005}). \label{noma_5}
    \end{align}
\end{subequations}
Constraint (\ref{noma_2}) is still not convex, so we approximate it at the point $(p_{b}^{[n]},\sigma_{b}^{[n]})$ as below,
\begin{equation}\label{pb}
\frac{p_{b}|h_b|^2}{\sigma_{b}^{[n]}}-\frac{\sigma_{b}p_{b}^{[n]}|h_b|^2}{\left(\sigma_{b}^{[n]}\right)^2}+\frac{p_{b}^{[n]}|h_b|^2}{\sigma_{b}^{[n]}}\geq \rho_{b}.
\end{equation}
$q$ can be obtained by exhaustive search, so constraint (\ref{noma_005}) can be omitted.
Thus, for a given $q$, the problem can be transformed to
\begin{equation}
    \begin{aligned}
       \max_{\mathbf{p_s}, p_b, \rho_b, \sigma_b}  & t  \label{noma_problem1}  \\ 
      \mathrm{s.t.} \ \ & (\ref{noma_1}), (\ref{pb}), (\ref{noma_3})-(\ref{noma_5}).
    \end{aligned}
\end{equation}
Now the problem is convex and the proposed SCA algorithm is summarized in Algorithm \ref{alg:NOMA}, where $\tau$ is the tolerance, so that the rate region of NOMA is obtained.

\begin{algorithm}[t!]
    \caption{SCA Algorithm for NOMA}\label{alg:NOMA}
    \textbf{Initialize}: $q=0$\;
    \Repeat{$q=N_{s}$}{\textbf{Initialize}:  $n\leftarrow0$, $t^{[n]}\leftarrow0$, $p_b^{[n]}$, $\sigma_b^{[n]}$\;
    \Repeat{$|t^{[n]}-t^{[n-1]}|<\tau$}{

    $n\leftarrow n+1$\;
    {Solve problem (\ref{noma_problem1}) using $p_b^{[n-1]}$ and $\sigma_b^{[n-1]}$, and denote the optimal value of the objective function as $t^{*}$ and the optimal solutions as $p_b^{*}$, $\sigma_b^{*}$}\;
    Update $t^{[n]}\leftarrow t^*$, $p_b^{[n]}\leftarrow p_b^{*}$, $\sigma_b^{[n]}\leftarrow\sigma_b^{*}$\;
    }}
    
\end{algorithm}
Since the solution of SCA algorithm at iteration $[n]$ is also a feasible solution in iteration $[n+1]$, the objective $t$ is non-decreasing. Therefore, the convergence of Algorithm \ref{alg:NOMA} is guaranteed. In the outer loop of  Algorithm \ref{alg:NOMA}, the computational complexity is $\mathcal{O}\left( N_s\right)$. For the inner loop, the number of iterations needed for convergence is $\mathcal{O} \left( \log \left(\tau^{-1} \right)\right)$. For each iteration, the computational complexity is $\mathcal{O}\left[N_s \right]^{3.5}$ \cite{9650662}. Hence, the computational complexity of Algorithm \ref{alg:NOMA} is $\mathcal{O}\left( \left[N_s\right]^{4.5}\log \left(\tau^{-1} \right)\right)$.

\subsection{RSMA}

In the uplink RSMA, a user can split its message into multiple parts instead of sending the whole message, and then it encodes them into multiple streams and sends the superposition of these streams \cite{485709}. Therefore, the first step of RSMA design is deciding which users will split the messages. In a conventional bit communication system, if there are $N$ users, splitting $N-1$ users can ensure that every point on the rate region is achievable \cite{485709}. However, for the coexistence of semantic users and a bit user, will splitting semantic users contribute to a better performance? From the above discussion in Section \ref{sec:2}, we know that the sentence similarity and the semantic rate are related to SINR, but splitting the semantic message will decrease the SINR. For example, we assume that a semantic user splits its message into two parts and encodes them into two streams, and the streams decoded first and decoded second are allocated power $p_{s,1,1}$ and $p_{s,1,2}$, respectively, where $p_{s,1,1}+p_{s,1,2}\leq p_{s,1}$. Thus, the SINR of the first stream is   
\begin{equation}
\rho_{s,1,1}=\frac{p_{s,1,1}|h_{s,1}|^2}{I_{s,1}+p_{s,1,2}|h_{s,1}|^2},
\end{equation}
where $I_{s,1}$ is the interference from other users and noise. If this semantic user does not split the message, its SINR is 
\begin{equation}
\rho_{s,1}=\frac{p_{s,1}|h_{s,1}|^2}{I_{s,1}} \geq \rho_{s,1,1}.
\end{equation}
Since this SINR decrease will deteriorate the sentence similarity, we do not split semantic users in the RSMA scheme.

\begin{figure}[]
\centering
\includegraphics[width=0.4\textwidth]{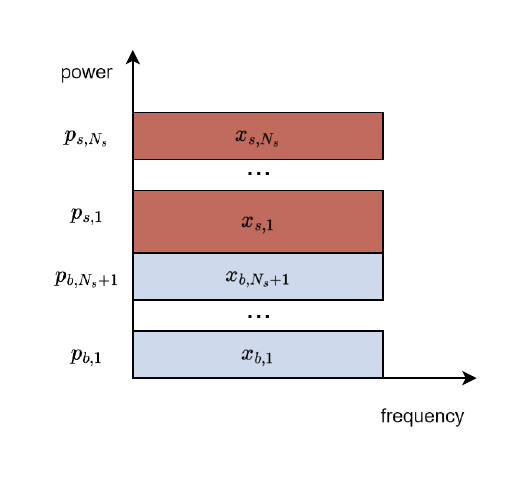}
\caption{The illustration of RSMA.}\label{fig:noma_decoding}
\end{figure}

\begin{figure}[]
\centering
\hspace{-1.5cm}\includegraphics[scale=0.6]{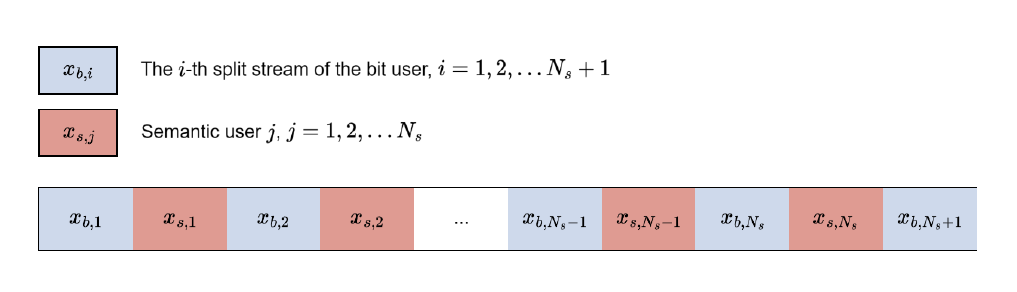}
\caption{The decoding order of RSMA.}\label{fig:rsma_scheme}
\end{figure}

After deciding on splitting the bit user only, we need to design the number of the split parts and the decoding order. Although in bit communication systems bit users usually split their messages into two parts \cite{9257190,9970313}, for a coexistence scenario with $N_s$ semantic users, it is more general for a bit user to split its message into $N_s+1$ parts and an illustration is shown in Fig. \ref{fig:rsma_scheme}. The semantic streams are denoted with red, and the split streams of the bit user are denoted with blue. Without the loss of generality, we assume that $x_{b,i}$ is the stream decoded $i$-th, i.e., $x_{b,1}$ is decoded the first and $x_{b,N_s+1}$ is decoded the last. There are two motivations of splitting a message into $N_s+1$ parts. First, $N_s+1$ is the maximum number of splitting parts and it provides the most flexibility in the decoding procedure. Let us take a toy example. We assume that we aim to find the highest achievable bit rate with a given semantic rate and sentence similarity. Intuitively, the optimal solution is allocating all the power to $x_{b,N_s+1}$. But it is not always possible because semantic users may not tolerate much interference from the bit user, and since we do not split the semantic users, we cannot obtain any flexibility by adjusting the decoding order of the split streams of semantic users. Therefore, we can allocate part of the power to $x_{b,N_s}$ to let part of the bit stream be decoded. If the semantic users still cannot tolerate the interference, we can allocate part of the power to $x_{b,N_s-1}$. In this way, at most $N_s+1$ streams may be allocated to power. The second motivation is that adjusting the power allocation of each stream can convert $N_s+1$ split streams to an arbitrary number of split streams between $1$ and $N_s$. For example, if we only allocate power to two streams and set the power of other streams as $0$, then it is equivalent to having two split streams. Another advantage is that we do not need to design the decoding order. If we set the number of split streams as two, there will be ${2 \choose N_s+1}$ decoding orders. However, if we generally set there are $N_s+1$ split streams and perform the optimization, we do not need to consider the decoding order. If two splitting streams is the optimal solution, in the optimization result only two streams will be allocated power. Therefore, the RSMA scheme illustrated in Fig. \ref{fig:rsma_scheme} is the most general design for the coexistence scenario. The received signal can be represented as 
\begin{equation}
y=\sum_{j \in \mathcal{J}} \sqrt{p_{s,j}} h_{s,j} x_{s,j}+ \sum_{k \in \mathcal{K}}\sqrt{p_{b,k}} h_b x_{b,k} +n,
\end{equation}
where $\mathcal{K}=\{1,2,..., N_s+1\}$, and $p_{b,k}$ is the power allocated to $x_{b,k}$. Thus, for $x_{b,k}$ the SINR is 
\begin{equation}
\rho_{b,k}=\frac{p_{b,k}|h_b|^2}{\sigma_n^2+\sum_{i=k}^{N_s}p_{s,i}|h_{s,i}|^2+\sum_{l=k+1}^{N_s+1}p_{b,l}|h_b|^2},
\end{equation}
and the rate of this stream is 
\begin{equation}
r_{b,k}=w\log_2 \left(1+\rho_{b,k}\right),
\end{equation}
so the achievable rate of the bit user is
\begin{equation}
r_b^R=\sum_{k=1}^{N_s+1} r_{b,k}.
\end{equation}
For semantic user $U_{s,j}$, the SINR is
\begin{equation}
    \rho_{s,j}=\frac{p_{s,j} |h_{s,j}|^2}{\sigma_{s,j}},
\end{equation}
where 
\begin{equation}
\sigma_{s,j}=\sum_{l=j+1}^{N_s+1}p_{b,l}|h_b|^2+\sum_{i=j+1}^{N_s}p_{s,i}|h_{s,i}|^2+\sigma_n^2, \ \forall j\in\mathcal{J}.
\end{equation}
Therefore, the achievable semantic rate is 
\begin{equation}
S_j^R=\frac{wI}{KL}\varepsilon (\rho_{s,j}).
\end{equation}
The rate region can be represented as 
\begin{equation}
\begin{aligned}
    C^{R} \triangleq \bigcup  \Bigg\{ &\left. \left(S^R, R_b^R\right): S^R\leq S_j^R, R_b^R \leq r_b^R, \varepsilon(\rho_{s,j})\geq S^{th}, \right. \\ & j\in \mathcal{J}, p_{s,j}\leq P, \sum_{k=1}^{N_s+1}p_{b,k} \leq P \Bigg\}.
    \end{aligned}
\end{equation}

For a given semantic rate $S$ inside the rate region, a problem of maximizing the bit rate can be formulated as below:

\begin{subequations} 
\begin{align}
\max_{\mathbf{p_s},\mathbf{p_b}} \  \ & r_b^R,  \label{problem00} \\
 \mathrm{s.t.} \ 
& \varepsilon \left(\frac{p_{s,j}|h_{s,j}|^2}{\sigma_{s,j}}\right)\geq S^{th},  \ \ \forall j\in \mathcal{J}, \label{p00_1}\\
& \frac{wI}{KL}\varepsilon \left(\frac{p_{s,j}|h_{s,j}|^2}{\sigma_{s,j}}\right)\geq S, \ \ \forall j\in\mathcal{J},  \label{p00_2}\\
& \sum_{k\in\mathcal{K}}p_{b,k}\leq P,  \label{p00_3}\\
& p_{s,j} \leq P,\ \ \forall j\in\mathcal{J},    \label{p00_4}
\end{align}
\end{subequations}
where $\mathbf{p_s}=\left[p_{s,1},..., p_{s,N_s}\right]$, $\mathbf{p_b}=\left[p_{b,1},..., p_{b,N_s+1}\right]$. Since (\ref{problem00}) is not convex due to its expression, we introduce two auxiliary variables $\boldsymbol{\rho_b}=\left[\rho_{b,1},..., \rho_{b,N_s+1}\right]$, $\boldsymbol{\sigma_b}=\left[\sigma_{b,1},..., \sigma_{b,N_s+1}\right]$, and we substitute (\ref{SNR_threshold}) and (\ref{semantic_rate_sinr}) into (\ref{p00_1}) and (\ref{p00_2}), respectively, and then the problem can be transformed to 
\begin{subequations} 
\begin{align}
\max_{\mathbf{p_s},\mathbf{p_b},\boldsymbol{\rho_b},\boldsymbol{\sigma_b}}  &\sum_{k\in \mathcal{K}} t_k,  \label{problem0} \\
 \mathrm{s.t.} \
& w\log_2 \left(1+\rho_{b,k}\right) \geq t_k, \ \forall k\in\mathcal{K},\label{p0_1}\\
\ & \frac{p_{b,k}|h_b|^2}{\sigma_{b,k}} \geq \rho_{b,k}, \ \forall k\in\mathcal{K},  \label{p0_2} \\
& \sigma_{b,k}\geq \sigma_n^2+\sum_{i=k}^{N_s}p_{s,i}|h_{s,i}|^2+\sum_{l=k+1}^{N_s+1}p_{b,l}|h_b|^2, \forall k\in\mathcal{K}, \label{p0_3} \\
& \frac{p_{s,j}|h_{s,j}|^2}{\sigma_{s,j}}\geq \gamma,  \ \forall j\in \mathcal{J}, \label{p0_4}\\
& \frac{p_{s,j}|h_{s,j}|^2}{\sigma_{s,j}}\geq \gamma_{sem}, \ \forall j\in\mathcal{J},  \label{p0_5}\\
& \sum_{k\in\mathcal{K}}p_{b,k}\leq P, \ \forall k\in\mathcal{K}  \label{p0_6}\\
& p_{s,j} \leq P, \forall j\in\mathcal{J}.    \label{p0_7}
\end{align}
\end{subequations}
Constraint (\ref{p0_2}) is still not convex, so we approximate it at the point $(p_{b,k}^{[n]},\sigma_{b,k}^{[n]})$ as below,
\begin{equation}\label{4}
\frac{p_{b,k}|h_b|^2}{\sigma_{b,k}^{[n]}}-\frac{\sigma_{b,k}p_{b,k}^{[n]}|h_b|^2}{\left(\sigma_{b,k}^{[n]}\right)^2}+\frac{p_{b,k}^{[n]}|h_b|^2}{\sigma_{b,k}^{[n]}}\geq \rho_{b,k}.
\end{equation}
Then the problem is transformed to 
\begin{equation} 
\begin{aligned}
\max_{\mathbf{p_s},\mathbf{p_b},\boldsymbol{\rho_b},\boldsymbol{\sigma_b}}  &\sum_{k\in \mathcal{K}} t_k, \label{problem000} \\
\mathrm{s.t.} \ \ &  (\ref{p0_1}), (\ref{4}) \ \mathrm{and} \ (\ref{p0_3})-(\ref{p0_7}).
\end{aligned}
\end{equation}
We can obtain the rate region by solving this convex problem and the above algorithm can be summarized as Algorithm \ref{alg:RSMA}.
\begin{algorithm}[t!]
    \caption{SCA Algorithm for RSMA}\label{alg:RSMA}
    \textbf{Initialize}:  $n\leftarrow0$, $t^{[n]}\leftarrow0$, $\boldsymbol{p_b^{[n]}}$, $\boldsymbol{\sigma_b^{[n]}}$\;
    \Repeat{$|t^{[n]}-t^{[n-1]}|<\tau$}{

    $n\leftarrow n+1$\;
    {Solve problem (\ref{problem000}) using $\boldsymbol{p_b^{[n-1]}}$ and $\boldsymbol{\sigma_b^{[n-1]}}$, and denote the optimal value of the objective function as $t^{*}$ and the optimal solutions as $\boldsymbol{p_b^{*}}$, $\boldsymbol{\sigma_b^{*}}$}\;
    Update $t^{[n]}\leftarrow t^*$, $\boldsymbol{p_b^{[n]}}\leftarrow \boldsymbol{p_b^{*}}$, $\boldsymbol{\sigma_b^{[n]}}\leftarrow \boldsymbol{\sigma_b^{*}}$\;
    }
\end{algorithm}

Algorithm \ref{alg:RSMA} is converged because the solution of SCA algorithm at iteration $[n]$ is also a feasible solution in iteration $[n+1]$, which means the objective $t$ is non-decreasing. The number of iterations needed for convergence is $\mathcal{O} \left( \log \left(\tau^{-1} \right)\right)$. In each iteration, the computational complexity is $\mathcal{O}\left[N_s \right]^{3.5}$. Hence, the computational complexity of Algorithm \ref{alg:NOMA} is $\mathcal{O}\left( \left[N_s\right]^{3.5}\log \left(\tau^{-1} \right)\right)$.

\section{Numerical Results}\label{sec:4}
In this section, we present the numerical results of the performances of different MA techniques in the coexistence scenario. The small-scale fading of the channels between the BS and all users is Rayleigh fading. The pass loss is modeled as $\rho=\rho_0 (1/d)^\beta$, where $\rho_0=-30$ dB
denotes the reference path loss at 1 meter, and $d$ is the distance between the BS and the user in meters. The available bandwidth $w=1 \ \mathrm{MHz}$, and $N_0=-140 \ \mathrm{dBm/Hz}$, so that $\sigma_n^2$ over the whole available bandwidth is $-80 \ \mathrm{dBm}$. Here we assume that the distances between the BS and the users are $30$ meters and $S^{th}=0.8$.
\subsection{The rate regions for different numbers of semantic users}
\begin{figure}
    \centering
    \includegraphics[width=0.45\textwidth]{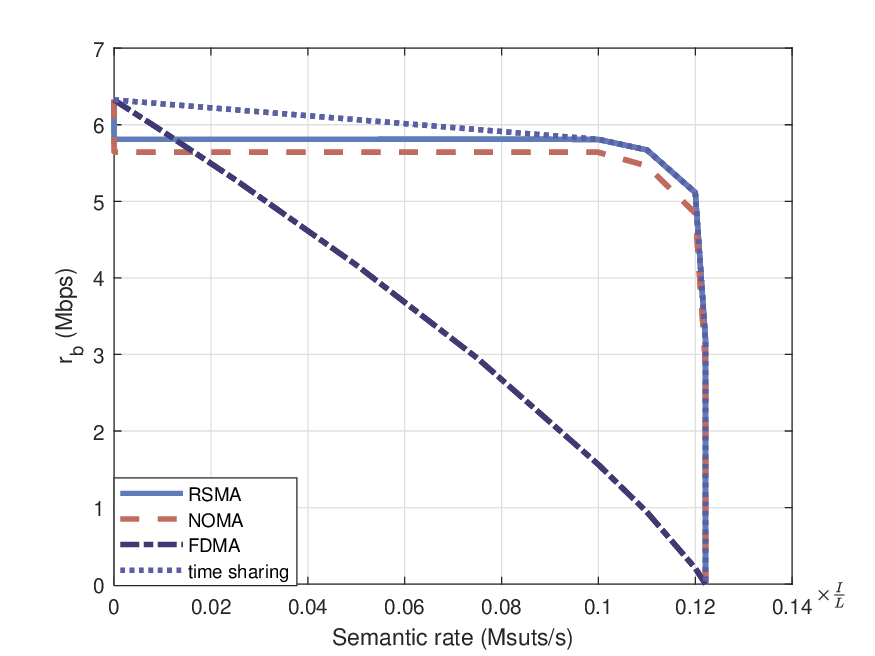}
    \caption{The rate region of the coexistence of one semantic user and one bit user.}
    \label{fig:two_user}
\end{figure}

    Firstly, let us start from a two-user case to have some intuition, which is one semantic user and one bit user and $K=8$. The rate region is shown in Fig. \ref{fig:two_user}, and it is very different from the rate region of a two-bit-user case. For FDMA, the relation between semantic rate and bit rate is approximately linear, which is intuitive, because for both semantic user and bit user the rate increases as bandwidth increases. However, for NOMA and RSMA, the rate regions are like squares. When the semantic rate is $0$, the semantic user does not transmit anything and the bit user will use all the resources to transmit, so the bit rate of NOMA and RSMA is the same as the one of FDMA. Then if the semantic user starts to transmit, $r_b$ drops suddenly. Recall the definition of sentence similarity and the semantic rate in (\ref{def_semilarity}) and (\ref{semantic_rate}), respectively. The semantic rate is decided by sentence similarity, while that sentence similarity must be higher than $S^{th}$ so that the semantic message can be decoded. Therefore, even if the semantic user does not require a high semantic rate, it still needs to achieve a high SINR to ensure sentence similarity. For example, for a given source of data, the semantic user $U_{s,1}$ requires $0.001 \ \mathrm{Msuts/s} \times \frac{I}{L}$, which gives that $\varepsilon(\rho_{s,1}) \geq 0.001K=0.008$, but only if $\varepsilon(\rho_{s,1})\geq S^{th}=0.8$ the message can be decoded, so that $\rho_{s,1}$ must achieve a specific level. Since both NOMA and RSMA use SIC, the user decoded first will tolerate interference from the other user, so that $r_b$ dropped suddenly to make sure that the semantic user can achieve a specific level of SINR. While FDMA allocates isolated bands to the users and they do not interfere with each other, so the relation between semantic rate and bit rate is approximately linear. Then, the bit rate of NOMA and RSMA keep the same level until the semantic rate increases to $0.1 \ \mathrm{Msuts/s}\times \frac{I}{L}$. When the semantic rate is lower than $0.1 \ \mathrm{Msuts/s}\times \frac{I}{L}$, the sentence similarity requirement is more stringent, which means as long as $\varepsilon(\rho_{s,1})\geq S^{th}$ any semantic rate below $0.1 \ \mathrm{Msuts/s}\times \frac{I}{L}$ can be achieved. When the semantic rate is higher than $0.1 \ \mathrm{Msuts/s}\times \frac{I}{L}$, the bit rates of NOMA and RSMA drop again because the semantic rate requirement is more stringent. Due to the message splitting of RSMA, RSMA has more flexibility and achieves a larger rate region. RSMA does not outperform FDMA when the semantic rate is below $0.01 \ \mathrm{Msuts/s}\times \frac{I}{L}$, because the bit user needs to sacrifice its SINR to ensure the sentence similarity of semantic users, while in FDMA the users do not interfere with each other. Nevertheless, RSMA outperforms FDMA in most situations. Besides, we can perform time-sharing between FDMA and RSMA to obtain the largest rate region, which is shown by the dotted line in Fig. \ref{fig:two_user}.

\subsection{Comparing RSMA behaviour for bit communications and for coexistence}
The main difference between RSMA scheme designs in conventional bit communications and in the coexistence scenarios is originated from the unique features of semantic communications. The effect of taking semantic features into account is intuitive if we look into the power allocation strategies. Let us still consider the two-user case to have some intuition. For the coexistence scenario, a two-user case is one bit user and one semantic user, and the bit user splits its message into two parts and encodes them into two streams. Thus, $\mathcal{K} \in \{1,2\}$ and the power allocated to the two streams are $p_{b,1}$ and $p_{b,2}$, and we use $\alpha=\frac{p_{b,1}}{p_{b,1}+p_{b,2}}$ to denote the percentage of the power allocated to $x_{b,1}$. After we solve (\ref{problem000}), we can obtain $\alpha$. For a conventional two-bit-user case, only one user will split its message, and we assume that user 1 transmits $x_{1,1}$ and $x_{1,2}$ with power $p_{1,1}$ and $p_{1,2}$, respectively. User 2 does not split the message and it transmits $x_{2}$ with power $p_2$. BS decodes the streams following the order of $x_{1,1}$, $x_2$ and $x_{1,2}$. We let $h_1$ and $h_2$ denote the channel for user 1 and user 2, respectively. For a given achievable rate of user 2, $R_2$, we can find the maximum achievable rate of user 1. Therefore, the problem can be formulated as 
\begin{subequations} 
\begin{align}
\max_{\substack{p_2,p_{1,1}, p_{1,2}, \\ \rho_{1,1},\sigma_{1,1}}}  &\sum_{k\in \{1,2\}} t_k,  \label{problem_bit} \\
 \mathrm{s.t.} \
& w\log_2 \left(1+\rho_{1,k}\right) \geq t_k, \ \forall k\in\{1,2\},\label{bit_1}\\
\ & \frac{p_{1,1}|h_1|^2}{\sigma_{1,1}} \geq \rho_{1,1}\ ,  \label{bit_2} \\
& \sigma_{1,1}\geq \sigma_n^2+p_{1,2}|h_{1}|^2+p_{2}|h_2|^2, \label{bit_3} \\
& \frac{p_{1,2}|h_{1}|^2}{\sigma_{n}^2}\geq \rho_{1,2}, \label{bit_4}\\
& p_2|h_2|^2\geq \left(2^{R_2}-1 \right)\left(\sigma_n^2+p_{1,2}|h_2|^2 \right),  \label{bit_5}\\
& p_{1,1}+p_{1,2}\leq P, \    p_{2} \leq P.\label{bit_6}
\end{align}
\end{subequations}
Then we can obtain the percentage of $p_{1,1}$, $\alpha=\frac{p_{1,1}}{p_{1,1}+p_{1,2}}$, by solving (\ref{problem_bit}).

\begin{figure}   
\centering
     \begin{subfigure}[b]{0.48\textwidth}         
         \includegraphics[scale=0.45]{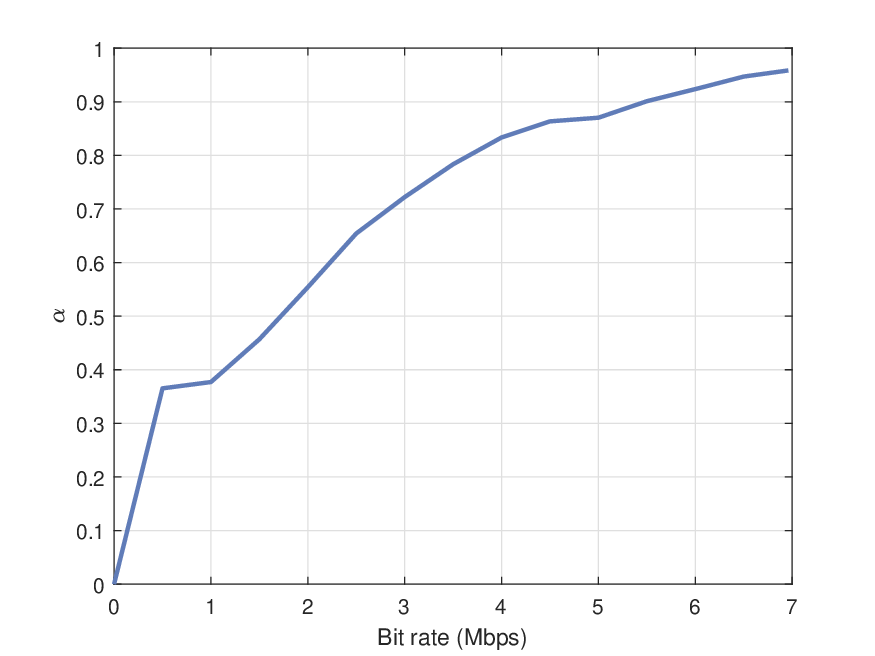}
         \caption{Two-bit-user case}
         \label{fig:bit_bit}
     \end{subfigure}     
     \begin{subfigure}[b]{0.48\textwidth}
         \includegraphics[scale=0.45]{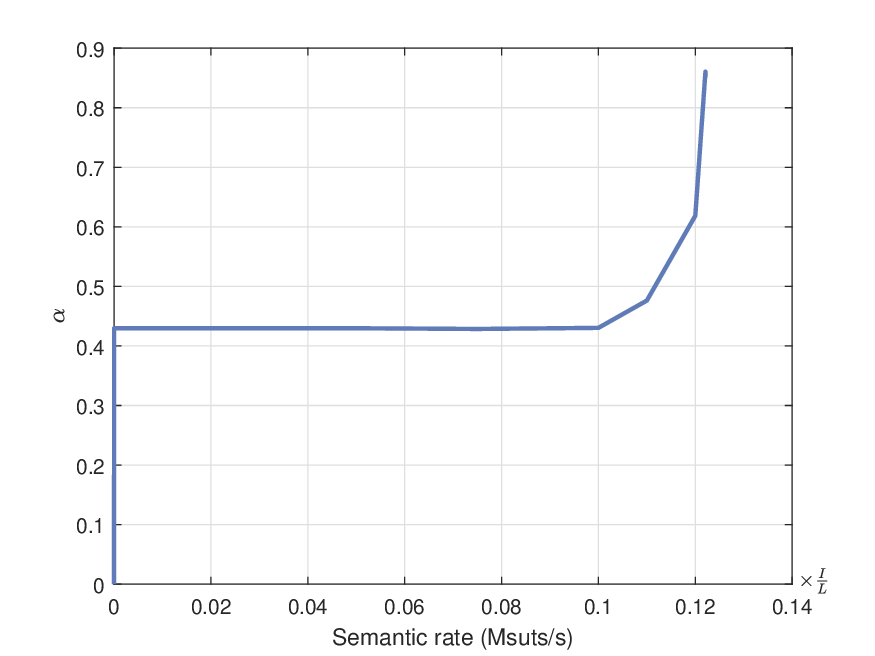}
         \caption{Coexistence case}
         \label{fig:bit_sem}
     \end{subfigure}   
    \caption{Power allocation strategies}
    \label{fig:split}
\end{figure}

Fig. \ref{fig:split} shows different power allocation strategies in two-bit-user case and coexistence case. Fig. \ref{fig:bit_bit} shows the relation between $\alpha$ and $R_2$. As $R_2$ increases, $\alpha$ increases. It means that to achieve a higher $R_2$, more power is allocated to $x_{1,1}$ so user 2 tolerates less interference. Fig. \ref{fig:bit_sem} shows the relation between $\alpha$ and semantic rate. When the semantic rate is 0, the semantic user does not transmit anything, so the bit user does not need to split, corresponding to $\alpha=0$. Once the semantic user starts to transmit, it should achieve a specific SINR to ensure sentence similarity, so $\alpha$ achieves a value around 0.42. Then $\alpha$ remains unchanged until the semantic rate reaches $0.1 \ \mathrm{Msuts/s} \times \frac{I}{L}$. These changes of $\alpha$ are also reflected in the rate region. The bit rate drops suddenly when semantic users start to transmit, and then it maintains the same till $0.1 \ \mathrm{Msuts/s} \times \frac{I}{L}$.

\subsection{The advantages of RSMA with multiple semantic users}
\begin{figure}
    \centering
    \includegraphics[width=0.45\textwidth]{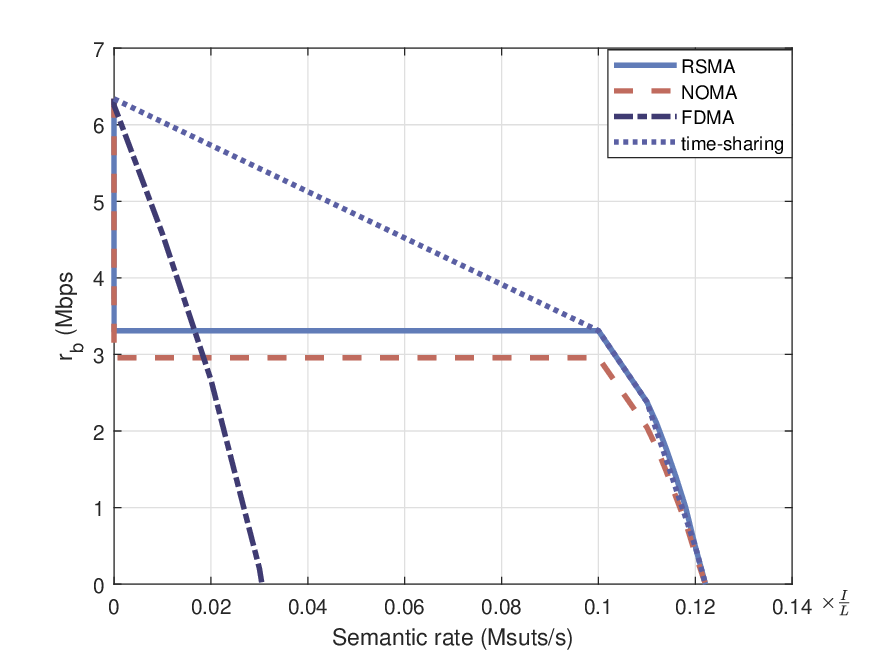}
    \caption{The rate region of the coexistence of four semantic users and one bit user.}
    \label{fig:four-user}
\end{figure}

The superiority of RSMA is more obvious when the number of semantic users increases. The rate regions of FDMA, NOMA and RSMA with four semantic users are shown in Fig. \ref{fig:four-user}. The rate regions are similar to the ones in Fig. \ref{fig:two_user}. For FDMA, the relation between the semantic rate and the bit rate is still approximately linear, but the rate region shrinks. Since FDMA allocates isolated resources to the users, the allocated bandwidth of each user is limited for a four-semantic-user case, so it is challenging to accommodate one bit user and four semantic users with a high semantic rate. Therefore, the bit rate drops to $0$ when the semantic rate is around $0.03 \ \mathrm{Msuts/s}\times \frac{I}{L}$. For NOMA and RSMA, their rate regions have similar shapes to the ones in Fig. \ref{fig:two_user}, but the bit rates of NOMA and RSMA drop more rapidly when the semantic users start to transmit, and they drop more slowly when the semantic rate is high. Since there are four semantic users, each user should achieve a certain SINR level to make sure its message can be decoded. This causes the bit user to tolerate more interference than the one-semantic-user case, so that the achievable bit rate decreases more when the semantic users start to transmit. Then the bit rates of NOMA and RSMA remain the same as they are in the one-semantic-user case. After the semantic rate is higher than $0.1 \ \mathrm{Msuts/s}\times \frac{I}{L}$ they drop again, but more slowly compared to Fig. \ref{fig:two_user}. In Fig. \ref{fig:two_user}, the bit rates almost drop vertically from $0.12 \ \mathrm{Msuts/s}\times \frac{I}{L}$. Recall the approximation of sentence similarity shown in Fig. \ref{fig:SNR_Similarity}, in the high $\rho$ regime, increasing $\rho$ hardly improves the sentence similarity, so the semantic rate increases very slowly. Therefore, when there are only one semantic user, it can achieve a high semantic rate easily, but it is very difficult to increase the semantic rate further. Even if a small increase in semantic rate requires a huge increase in SINR. Thus, the bit rate drops almost vertically. While for a four-semantic-user case, it is not easy to achieve a high semantic rate. The semantic rate increases as the transmission power increases, so the bit user needs to tolerate more interference from semantic users and the bit rate decreases gradually from $0.1 \ \mathrm{Msuts/s}\times \frac{I}{L}$. Compared to the one-semantic-user case, the advantage of RSMA is more manifest. In the four-semantic-user case, the bit user splits its message into five parts, and it provides more flexibility to the SIC procedure compared to NOMA. For example, in NOMA the bit user may be the first to be decoded so that the semantic users can be decoded successfully. Thus, the bit rate is low because the bit user tolerates the interference from four semantic users. In RSMA, BS decodes split streams of bit user and semantic streams alternatively. Thus, the split streams do not always need to tolerate high interference, so a higher bit rate can be achieved. Similarly, the largest rate region can be achieved by performing time sharing between FDMA and RSMA, which is denoted by the dotted line. We also observe that in most instances, splitting two streams can achieve the best performance, but sometimes the optimal result is obtained by splitting a message more than two times. Besides, we observe that when splitting two streams is the optimal solution, the decoding order is not always the same. Therefore, considering the most general case, splitting $N_s+1$ streams, would skip the decoding order design and reduce the complexity in the optimization.  

\begin{figure}
    \centering
    \includegraphics[width=0.45\textwidth]{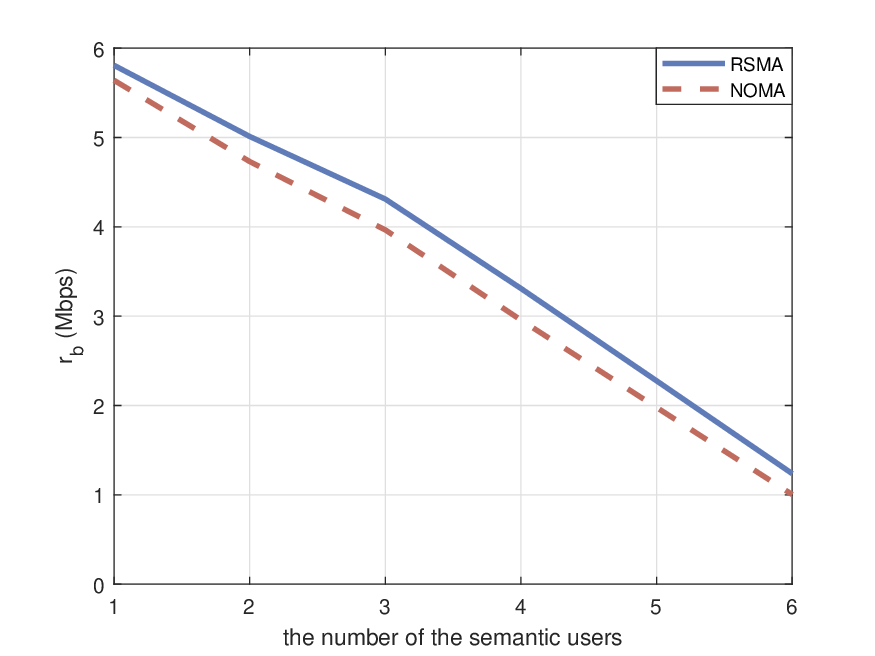}
    \caption{The number of semantic users versus the achievable bit rate.}
    \label{fig:Bit_vs_number}
\end{figure}

The relation between the number of semantic users and the achievable bit rate is shown in Fig. \ref{fig:Bit_vs_number}. We still set $S^{th}=0.8$ and $K=8$, and we fix the semantic rate at $0.1 \ \mathrm{Msuts/s}\times \frac{I}{L}$. We do not present FDMA because the semantic users cannot always achieve $0.1 \ \mathrm{Msuts/s}\times \frac{I}{L}$ in FDMA. For both NOMA and RSMA, the relations between the number of semantic users and achievable bit rate are roughly linear. There is a constant gap between RSMA and NOMA, which shows as the number of semantic users increases the superiority of RSMA is more obvious. The reason is that RSMA allows the bit user to split its message into multiple parts, so it brings more flexibility during the decoding procedure and the bit user can achieve a higher bit rate.

\subsection{The advantages of RSMA in high accuracy applications}
\begin{figure}
    \centering
    \includegraphics[width=0.45\textwidth]{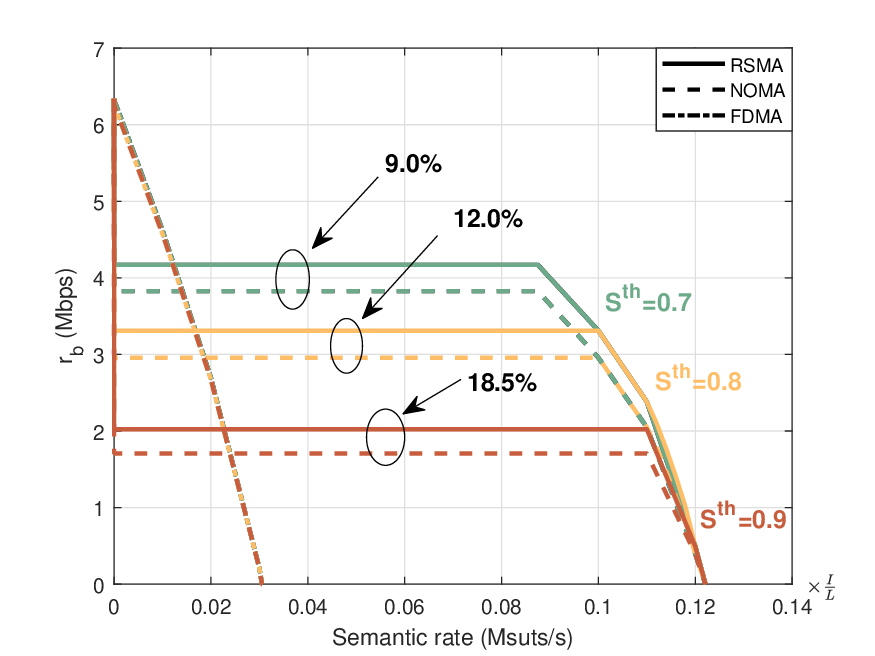}
    \caption{The rate regions of $S^{th}=0.7, 0.8, 0.9$, $N_s=4$.}
    \label{multipleth}
\end{figure}

The rate regions with different $S^{th}$ are shown in Fig. \ref{multipleth}. $S^{th}=0.7, 0.8, 0.9$ are denoted by green, yellow and red, respectively. RSMA, NOMA and FDMA are represented by solid line, dash line, and dash-dotted line, respectively. For FDMA, the rate regions of different $S^{th}$ are almost the same. From Fig. \ref{fig:SNR_Similarity}, we know that the sentence similarity approaches the upper bound when SNR achieves a specific level, which means the sentence similarity hardly improves even if increasing SNR. Since the users do not interfere with each other, the SINR of each user is relatively high and their sentence similarities are very close to the upper bound. Therefore, $S^{th}$ does not affect the resource allocation strategy. However, for NOMA and RSMA, higher $S^{th}$ means a higher SINR of each user, so the semantic users need to use more transmission power. Thus, the bit user needs to tolerate more interference from others and the rate regions become smaller as $S^{th}$ increases. Compared to NOMA, RSMA always has a larger rate region and this is benefited from message splitting. Fig. \ref{multipleth} shows that RSMA has more improvement with a higher $S^{th}$. When the requirement is more stringent, the flexibility of RSMA is more helpful. When $S^{th}=0.7$, the improvement of RSMA is $9.0 \%$, and while when $S^{th}=0.9$, the improvement of RSMA is $18.5 \%$. These results show that RSMA is more robust to a stringent sentence similarity requirement, which means it has the potential to be used in applications with high accuracy.

\section{Conclusion}\label{sec:5}
In this work, we investigate the coexistence of semantic users and a bit user in the uplink. Three MA techniques are studied for the coexistence, namely FDMA, NOMA and RSMA. Specifically, we propose a RSMA scheme for this coexistence scenario. We characterize the rate regions for FDMA, NOMA and RSMA to study their performances. For NOMA and RSMA, to obtain the boundary points on the rate regions, we propose SCA algorithms to transform a non-convex problem into a convex problem and then solve it. We simulate the rate regions with different numbers of semantic users and various $S^{th}$, and we find that RSMA always outperforms NOMA but not FDMA, but we can perform time sharing to obtain the optimal performance. We summarize the main changes in RSMA design and behaviours in the coexistence scenario: 1) RSMA does not split semantic users and it splits the bit user multiple times. 2) The rate regions of the coexistence scenario and power allocation strategies are quite different from the ones of bit communications. They are caused by considering the features of semantic communications. 3) Due to introducing the semantic users, RSMA cannot always outperform OMA, but the largest rate regions can be obtained by time sharing between RSMA and OMA. In conclusion, the promising results show that RSMA has the potential to be applied to the coexistence of semantic users and bit users in future networks.


%

\ifCLASSOPTIONcaptionsoff
  \newpage
\fi



%
 
\bibliographystyle{IEEEtran} 
\bibliography{references.bib} 

%




%



\end{document}